\documentstyle[prd,aps,floats,tabularx,epsfig]{revtex}
\newcommand{\be}{\begin{equation}}
\newcommand{\ee}{\end{equation}}
\newcommand{\bea}{\begin{eqnarray}}
\newcommand{\eea}{\end{eqnarray}}

\begin{document}
\twocolumn[\hsize\textwidth\columnwidth\hsize\csname@twocolumnfalse\endcsname
\title{Perturbation evolution with a non-minimally coupled scalar field}
\author{Rachel Bean,\\
Theoretical Physics, The Blackett Laboratory, \\
Imperial College, Prince Consort Rd.,\\
London SW7 2BZ}
\date{28th April 2001}
\maketitle

\begin{abstract}
We recently proposed a simple dilaton-derived quintessence model in which the scalar field was non-minimally coupled to
cold dark matter, but not to `visible' matter. Such couplings can be attributed to the dilaton in the low energy limit of
string theory, beyond tree level. In this paper we discuss the implications of such a model on structure formation, looking
at its impact on matter perturbations and CMB anisotropies. We find that the model only deviates from $\Lambda$CDM and
minimally coupled theories at late times, and is well fitted to current observational data. The signature left by the
coupling, when it breaks degeneracy at late times, presents a valuable opportunity to constrain non-minimal couplings
given the wealth of new observational data promised in the near future.

\medskip

\end{abstract}
\bigskip
]

\section{Introduction}

There is recent evidence (\cite{perl}-\cite{riess}) that the Universe's expansion is accelerating. If this is
so, it would have fundamental cosmological implications, for progressing the dark matter problem and reconciling a high
Hubble Constant, $h\sim 0.65$, with an old Universe $t_{0}>11Gyr$. To explain such an
acceleration, the Universe would have
to have a matter component, additional to ordinary matter and radiation, since the latter two have equations of state that are unable to
generate the required kinematics. In line with current observational constraints,  the additional matter would have to have
 an equation of state $p= w\rho $ with $w \in (-1,-0.4)($\cite{turn}-\cite{perl2}).

A pure cosmological constant cannot explain the observed acceleration without running into fine tuning problems; one would
need $\Lambda\sim 10^{-122} c^{3}/(\hbar G)$, several hundreds of orders of magnitude lower than one would expect from a
vacuum energy originating at the Planck time \cite{BT}. This has lead to a wealth of proposals using a scalar
``quintessence'' field, minimally coupled to matter through gravity,  which can be cajoled into acting as an effective
cosmological constant in the presence of a suitable potential. Models of particular interest use ``tracker'' potentials
(e.g.\cite{wett}-\cite{zlat})
which allow the scalar field to produce the required dynamics without dependence on initial conditions, but these still
require small-scale parameters. More recently, a model was proposed (\cite{albsko},\cite{albsko2}) with a potential who's
parameters were, a more physically agreeable, Planck scale.
Explaining why the acceleration has only arisen recently, however, still requires some degree of fine-tuning in the model
parameters, if not in the initial conditions, in order to confine acceleration to the current epoch \cite{us}. A more
practical explanation for the coincidental acceleration nowadays is that we are in close proximity to the cosmological
transition from radiation to dust domination. Armendariz-Picon , Mukhanov and Steinhardt \cite{kappa} utilised this
proximity to drive the dynamics of their $\kappa$-essence model although the Lagrangian used is somewhat complex,
consisting of a series of non-linear kinetic terms.

In a recent paper \cite{us2}, we proposed a simpler model which harnesses the dynamical shift in the radiation-dust
transition using a non-minimally coupled scalar field. We showed that a coupling of this form can use the transition to
dust domination to push a quintessence field off scaling behaviour, and produce acceleration in the background nowadays.

In this paper we consider the impact of such a non-minimal coupling on the evolution of perturbations to the background
and the subsequent implications for both structure formation and the Cosmic Microwave Background (CMB).

We start by giving an overview of the coupled quintessence model, and then go on to discuss the implications of coupling
for perturbation evolution and structure formation.

\section{Coupled quintessence model}

Non-minimal theories are commonly expressed in one of two frames. In one, the problem is posed in the Jordan frame and the
scalar field is directly coupled to curvature, in the form $f(\phi)\mathcal{R}$, and produces a departure from Einstein's
gravity, as is seen in Brans-Dicke theories \cite{bradic}. This effect was used by \cite{perr} to force the quintessence
field out of scaling behaviour, necessary to give accelerated dynamics, however this ``R-boost'' occurs early in the
radiation epoch and cannot explain acceleration today.
In the second, the Einstein frame is used and the scalar field instead couples to terms in the matter Lagrangian resulting
in dynamical, field-dependent, masses and polarisations. These two groups are interrelated through conformal transformation
 of the metric; any theory in one frame can be rephrased in the other. However, usually a simple function in one frame is mapped into a
 complicated function in the other. Such couplings are heavily constrained
when applied to the visible matter in the Universe, whether to
photons \cite{carroll}, or to what is usually called baryons \cite{dam94}.
However, it could be that the dilaton coupled differently to visible
matter and to the dark matter of the Universe. This hypothesis
was suggested in \cite{dam90}, and allows for large couplings
to be consistent with observations (see also \cite{amen},\cite{uzan}). We consider a scenario in which such a case exists.

We choose $g^{\mu\nu}$ to have convention (+ - - -) in a flat FRW background. All quantities are expressed in units with
$M_{P}=(8\pi G_{N})^{-1/2}=1$ where $M_{P}$ is the Planck mass and $G_{N}$ is the Newtonian Gravitational constant.
We consider a Lagrangian of the form:

\begin{equation}\label{action}
{\cal L}={\sqrt{-g}} {\left( -{R\over 2}
+{1\over 2}\partial_\mu\phi\partial^\mu\phi-V(\phi)
+{\cal L}_V+f(\phi){\cal L}_{c} \right)}
\end{equation}
in which ${\cal L}_V$ is the Lagrangian of ``visible matter''
(baryons, photons, and also baryonic and neutrino dark matter),
and ${\cal L}_{c}$ the Lagrangian of a dominant non-baryonic form
of cold dark matter.We take $V(\phi)=V_0e^{-\lambda\phi}$ the standard
quintessence potential, which drives scaling behaviour when the coupling is minimal (\cite{ferjoy1},\cite{ferjoy}).

The coupling investigated is of the form $f(\phi)=1+\alpha(\phi-\phi_{0})^{\beta}$. Couplings of this form could arise as
generalisations of an effective action for massless modes of a dilaton \cite{dam94} after performing a conformal
transformation from the string frame into the Einstein frame.
$\alpha$ and $\beta$ are parameters reflecting the shape of the minimum being approached by the coupling \cite{tsey}.

\subsection{Background Evolution}
\begin{figure}[ht]
\begin{center}
\epsfxsize=8.2cm \epsfysize=8.2cm \epsffile{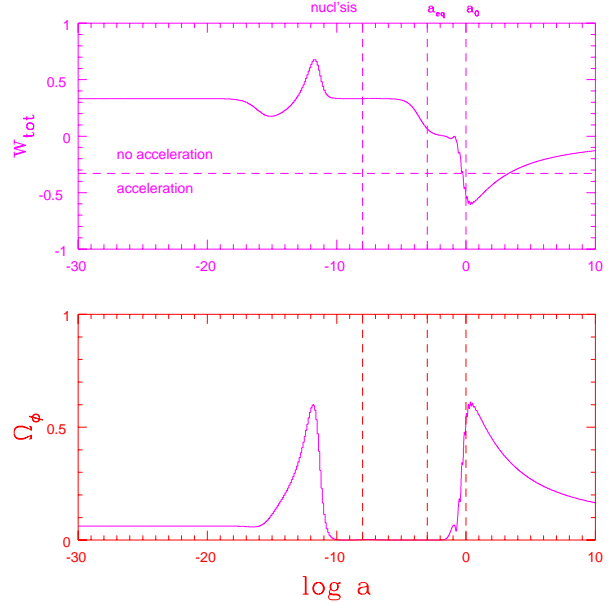} \caption{The evolution of $\Omega_\phi$ and
$w_{tot}$ for a model with $\lambda=8$, $\beta =8$ , $\alpha =50$, and $\phi_0=32.5$ (and
$\Omega_b=0.053, h=0.65)$. An early period of scaling is broken near the transition from radiation to
matter, first with a period of kination, then inflation. At late times the universe returns to a
matter dominated scaling solution.}
\end{center}
\label{bkg}
\end{figure}

\begin{figure}[ht]
\begin{center}
\epsfxsize=8.2cm \epsfysize=8.2cm \epsffile{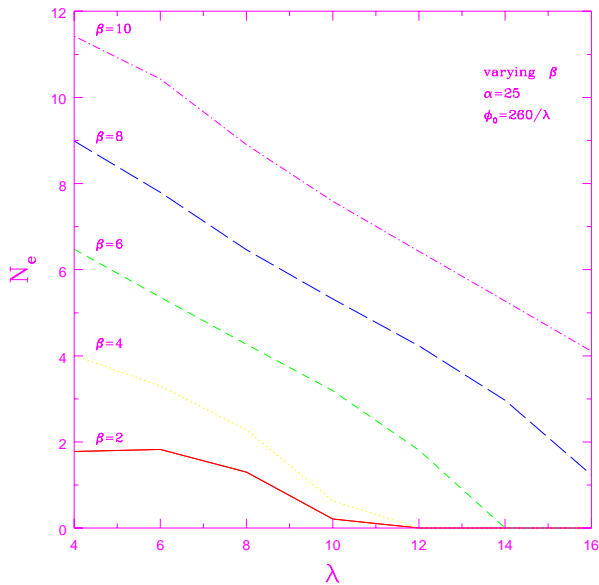} \caption{The amount of accelerated expansion
produced with the model for various values of $\beta$, measured in the number of e-foldings
$N_e=a_f/a_i$. $a_i$ and $a_f$ are the expansion scales when inflation begins and ends, respectively,
($a_i<a_0\le a_f$, where $a_{0}=1$ is the expansion scale nowadays). We have given $\alpha=25$ as an
example with $\phi_0=260/\lambda$.}
\end{center}
\label{fine1}
\end{figure}

\begin{figure}[ht]
\begin{center}
\epsfxsize=8.2cm \epsfysize=8.2cm \epsffile{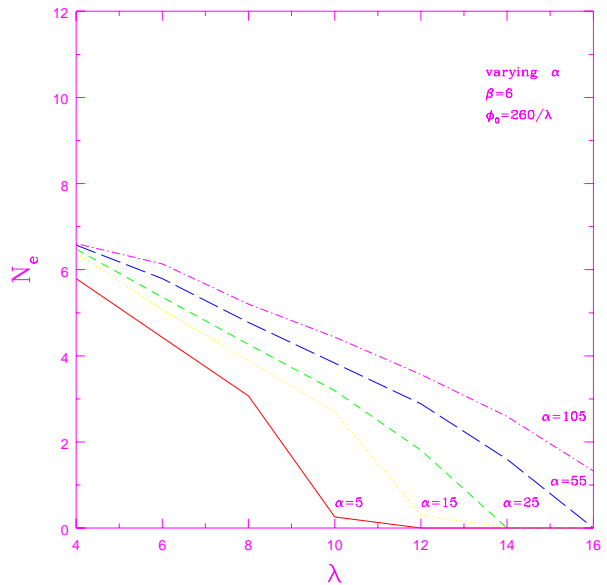} \caption{The amount of accelerated expansion
produced with the model as $\alpha$ is varied, measured again in the number of e-foldings $N_e$.We
have given $\beta=6$ as an example with $\phi_0=260/\lambda$.}
\end{center}
\label{fine2}
\end{figure}

Here we discuss the background equations in the conformal FRW metric, which are derived in appendix \ref{APP1}.
The field equations are obtained by varying the action with respect to the metric and the scalar field:
\begin{eqnarray}
G_{\mu\nu}&=&T^{(V)}_{\mu\nu}+T^{(\phi)}_{\mu\nu}+f(\phi)T^{(c)}_{\mu\nu}\\
\nabla^{2} \phi&=&{\partial V\over \partial \phi}-{\partial f
\over \partial \phi}{\cal L}_{c}
\end{eqnarray}
where $G_{\mu\nu}$ is the Einstein's tensor and the various $T_{\mu\nu}$
are stress-energy tensors.
Heuristically, we may interpret the new term driving $\phi$
as a contribution to an effective potential $V_{eff}=V-f(\phi){\cal L}_{c}$.
Bianchi's identity ($\nabla_\mu G^\mu_\nu=0$) leads to
:
\begin{eqnarray}
\nabla_\nu T^{\mu\nu(V)}&=&0\\
\nabla_\nu T^{\mu\nu(c)}&=& (g^{\mu\nu}{\cal L}_{c}
-T^{\mu\nu}_{c})
{f'\over f}\nabla _{\nu}\phi
\end{eqnarray}
These are to be contrasted with Amendola's coupled
quintessence \cite{amen1} (for which the interaction term is proportional to $T$).

Evaluating the components of the field equations, with scale
factor $a$, we find Friedmann equations:
\begin{eqnarray}
{3\over a^{2}}{\left(\dot a \over a\right)} ^2 &=&
\rho_{b}+\rho_{\gamma}+f(\phi)\rho_{c}+\frac{1}{2}{\dot{\phi}\over a}^{2}+V(\phi)
\\
\dot{\rho}_{c}+3\displaystyle{\frac{\dot{a}}{a}}\rho_{c}&=&-{f'(\phi)\dot{\phi}\over f(\phi)}
(\rho_{c}+{\cal L}_{c})=0 \label{cdmbk} \\ \rho_b+3\displaystyle{\frac{\dot{a}}{a}}\rho_{b}&=&0 \\
\rho_{\gamma}+4\displaystyle{\frac{\dot{a}}{a}}\rho_{\gamma}&=&0 \\
\ddot{\phi}+2\displaystyle{\frac{\dot{a}}{a}}\dot{\phi}+a^{2}V'&=&f'(\phi){\cal
L}_{c}a^{2}=-f'(\phi)\rho_{c}a^{2} \label{phieqn}
\end{eqnarray}
where dots represent derivatives with
respect to conformal time, and the prime (') indicates differentiation
with respect to $\phi$.

One notices in equation (\ref{cdmbk}) that the evolution of the background coupled dark matter is unaffected by the coupling. This simply arises because we are coupling to pressureless matter for which ${\cal L}_c=-\rho_c $; if we had instead coupled to radiation we would find ${\cal L}$=0 and the coupling would have altered the background evolution (as discussed in appendix \ref{APP1}). However as will be discussed later, observations measure the coupled energy density $f(\phi)\rho_{c}$ not simply $\rho_{c}$ so that the magnitude of the observed matter is affected by the coupling through (\ref{phieqn}).  

Fig.1. shows the evolution of $\Omega_{\phi}$ and overall equation of state $w_{tot}=\rho_{tot}/p_{tot}$ for one model
scenario. One can see that deep in the radiation epoch the coupling has a negligible effect on the overall dynamics and the
scalar field's energy density scales with that of the dominant radiation, as in the minimally-coupled case. As the
transition from radiation to matter domination is approached the coupling becomes important and the dynamics are driven
away from scaling behaviour. The driving term on the right hand side of equation (\ref{phieqn}) first, transiently, drives
the field to kinate, suppressing the evolution of the scalar field and $\Omega_{\phi}\sim 0$ , then it re-emerges into
inflationary behaviour to provide the accelerated expansion we observe today. The model requires that $\beta$ be even and
that the value of $\phi_{0}$ is of the order of magnitude of the scalar field today. However given these constraints, the model provides acceleration for a wide range of parameters as shown in the parameter space plots for the non-minimally coupled model
in Figs. 2 and 3.

In minimally coupled models with exponential potentials, the value of the parameter $\lambda$ is limited by BBN constraints
\cite{BHM} to be $\lambda\ge 8$ however the NMC model avoids this constraint through the suppression of $\Omega_\phi$ at
nucleosynthesis, irrespective of $\lambda$'s value. Parameter constraints for the non-minimal case can only therefore come from CMB and matter
power spectrum predictions discussed below.  In order to compare the non-minimal models with analogous minimally coupled ones however, we consider cases with $\lambda=8$ in our discussion below.

\subsection{Observational implications}

\begin{figure}[ht]
\begin{center}
\epsfxsize=8.2cm \epsfysize=8.2cm \epsffile{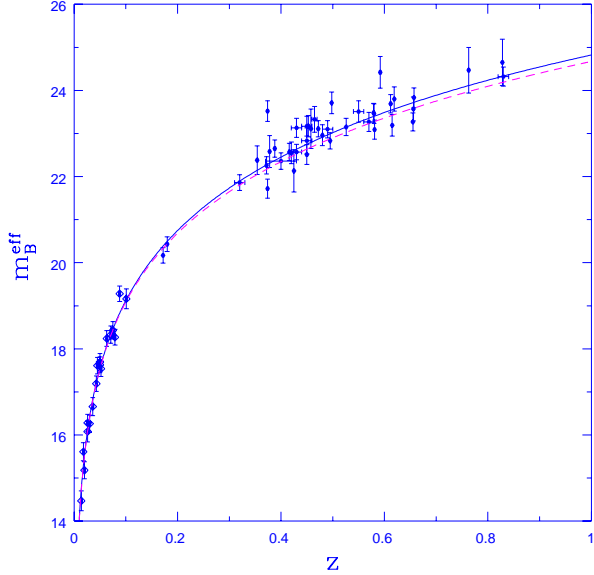} \caption{Plot of the effective bolumetric magnitude for the Cal\'{a}n Tololo (open diamonds) and SCP data points (solid circles) against redshift. The curves correspond to two models considered in this paper $\Lambda$ CDM model (solid line) with  $\Omega_c=0.347, \Omega_b=0.053$ and $\Omega_\Lambda=0.6$ and a non-minimally coupled model with  $f(\phi)\Omega_c=0.347, \Omega_b=0.053$ and $\Omega_\phi=0.6$ with model parameters specified in Fig.1.}
\end{center}
\label{hubble}
\end{figure}
It is pertinent to consider whether the effect of the non-minimal coupling on the background at late times could be seen in current observations, i.e. when looking at the predicted apparent magnitude versus redshift relation at $z<2$.  

The apparent bolumetric magnitude is given by
\begin{eqnarray}
m(z)=M+5\log d_{L}(z)+25 
\end{eqnarray}
where $M$ is the absolute bolumetric magnitude, and $d_{L}$ is the luminosity distance in Mpc 
\begin{eqnarray}
d_{L}=(1+z)\int_{0}^{z}{dz'\over H(z')}
\end{eqnarray}
In Fig.4 we plot the effective bolumetric magnitude from the B-band filter, $m^{eff}_{B}(z)$, from the Cal\'{a}n Tololo \cite{calan} and SCP \cite{perl2} surveys and predicted  m(z) curves for the non-minimally coupled model in Fig.1 and a comparative $\Lambda$ CDM model. The effective magnitude is obtained from the apparent magnitude after taking into account the lightcurve width-luminosity correction, galactic extinction and the K-correction from the differences in the observed R-band and restframe B-band filters \cite{perl2}.
Within current observational error constraints, the non-minimally coupled model cannot be distinguished from the $\Lambda$CDM model. Recently proposed observational projects (see for example \cite{jochalb}) may offer future hope to discriminate between the effect of quintessence models on the background evolution.

In the remainder of the paper, we consider an alternative approach to distinguishing between quintessence models, through their effect not on the background but on the perturbations about it. 

\section{Implications for structure formation}

The addition of the scalar field has implications for structure formation both due to the addition to the homogeneous background
 energy density, and secondly by the generation and evolution of scalar field perturbations. The additional
background energy density shifts the equality redshift and alters the angular distance to the last scattering surface. The
scalar field also introduces extra terms in the perturbed Einstein equations and opens up the possibility of isocurvature
perturbations evolving.

We study the impact of these effects by calculating the linear perturbation equations and specifying the initial conditions.
These are then evolved from early on in the radiation epoch when the coupling is unimportant through to nowadays.
The matter and CMB power spectra are then calculated and compared with those obtained with minimally coupled models
and observations.

\subsection{Linear perturbation evolution}  \label{lineq}

We follow the approach and notation of Ma and Bertschinger \cite{mabert} extended by Ferreira and Joyce \cite{ferjoy} for
minimally coupled scalar fields. A simplified model containing no baryons is used for the discussion, although a full
theory containing baryons and relativistic neutrinos is used to obtain the CMB and matter power spectrum predictions
presented.
The essential results are presented here, while a full derivation of the equations can be found in the appendix \ref{APP2}.

Consider perturbations to a flat FRW metric in the synchronous gauge, with line element
\begin{equation}
ds^{2}=a(\tau)^{2}\left\{-d\tau^{2}+(\delta_{ij}+h_{ij})dx^{i}dx^{j}\right\}
\end{equation}

We will only be concerned with the scalar modes of the perturbation, for which we can parameterise the metric perturbation
as
\begin{equation} h_{ij}=\int d^{3}ke^{ik.x}[\hat{k}_{i}\hat{k}_{j} h(k,\tau)+( \hat{k}_{i}\hat{k}_{j}-
\frac{1}{3}\delta_{ij}6\eta(k,\tau))]
\end{equation}
where h is the trace of the metric perturbation. To obtain the linear perturbation evolution equations we consider the
perturbed Einstein equations
\begin{eqnarray}
k^{2}\eta-\frac{1}{2}H\dot{h}&=&4\pi Ga^{2}\delta T_{0}^{0}
\\ k^{2}\dot{\eta}&=&4\pi Ga^{2}ik_{i}\delta T_{i}^{0}
\\ \ddot{h}+2H\dot{h}-2k^{2}\eta&=&-8\pi Ga^{2}\delta T_{i}^{i}
\\  \ddot{h}+6 \ddot{\eta}+2H(\dot{h}+6\dot{\eta})-2k^{2}\eta&=&24\pi Ga^{2}(\hat{k}_{i}\hat{k}_{j}-\frac{1}{3}\delta_{ij})
\Sigma^{i}_{j}
\end{eqnarray}
where $\Sigma^{i}_{j}$ is the traceless shear. Writing the perturbations to energy densities, $\rho$,
pressures, $p$,
and the scalar field , in terms
of a homogeneous background plus a perturbation, we have
\begin{eqnarray}
\rho(x,\tau)=\rho(\tau)(1+\delta(x,\tau))
\\ p(x,\tau)=p(\tau)+ \delta p(x,\tau)
\\ \Phi(x,\tau)=\phi(\tau)+\varphi(x,\tau)
\end{eqnarray}
The only perturbation in $T^{\mu}_{\nu}$ to be affected by the coupling is $\delta T_{0}^{0}$, the other perturbations are
the same as for a minimally coupled model,
\begin{eqnarray}
\delta T_{0}^{0}&=&-\rho_{\gamma}\delta_{\gamma}-(\varphi f'+f\delta_{c})\rho_{c}-(\frac{1}{a^{2}}\dot{\varphi}\dot{\phi}+
\varphi V')
\\ ik_{i}\delta T_{i}^{0}&=&\frac{4}{3}\rho_{\gamma}\theta_{\gamma}+\frac{1}{a^{2}}\dot{\phi}\nabla^{2}\varphi
\\ \delta T_{i}^{i}&=&3\left(\frac{1}{3}\rho_{\gamma}\delta_{\gamma}
+\frac{1}{a^{2}}\dot{\varphi}\dot{\phi}-\varphi V'\right)
\end{eqnarray}
where $\theta$ is the velocity divergence. The evolution equations of the density perturbations for radiation and the dark
matter component are the final requirement. One finds, as is shown in \ref{APP2}, that the coupling does not effect the first order equation for the matter perturbation so that,
\begin{eqnarray}
\delta_{\gamma}=-\frac{4}{3}\theta_{\gamma}-\frac{2}{3}\dot{h}
\\ \delta_{c}=-\theta_{c}-\frac{1}{2}\dot{h} \label{foe}
\end{eqnarray}
The spare degree of freedom in the synchronous gauge allows us to choose the background, synchronous coordinates. As is
conventional, we do this by constraining the dark matter field such that $\theta_{c}=0$, which fixes
$\dot{\delta_{c}}=-\frac{1}{2}\dot{h}$.
 We are now able to write down the
perturbation equations for the non-minimally coupled system.

\begin{eqnarray}
\ddot{\delta}_{c}&+&H\dot{\delta}_{c}+\frac{3H^{2}\Omega_{c}f}{2}\delta_{c}
 \nonumber \\ &=&-3H^{2}\Omega_{\gamma}\delta_{\gamma}-2\dot{\varphi}\dot{\phi}+(a^{2}V'-\frac{3H^{2}\Omega_{c}f'}{2})
 \varphi \label{pert}
\\ \ddot{\varphi}&+&2H\dot{\varphi}+[k^{2}+a^{2}(V''+f''\rho_{c})]\varphi  \nonumber\\ &=&-\frac{1}{2}\dot{h}\dot{\phi}-a^{2}f'\rho_c\delta_c \label{pert2}
\\ \ddot{\delta}_{\gamma}&-&\frac{k^{2}}{3}\delta_{\gamma}=\frac{4}{3}\ddot{\delta_{c}}
\end{eqnarray}

The non-minimal coupling introduces extra terms into the equations for matter and scalar field perturbations,
altering the mass terms and source terms , the latter shown on the right hand side of the equality for clarity.
The coupling will only affect the radiation perturbations indirectly through the background bulk (via H) and through
$\ddot{\delta_{c}}$.

Deep in the radiation epoch, the coupling to dark matter is
unimportant. The adiabatic perturbation evolution closely follows the
power-law solutions for the minimally coupled model with an
exponential potential as discussed by Ferreira and Joyce \cite{ferjoy}.
The growing modes of $\delta_{\gamma},\delta_{c}$ and
$\varphi$ evolve $\propto \tau^{2}$
\begin{eqnarray}
\delta_{\gamma}&=&-\frac{2}{3}C(k\tau)^{2}, \ \ \
\delta_{c}= -\frac{1}{2}h =\frac{3}{4}\delta_{\gamma}, \nonumber \\
 \varphi&=&-\frac{2}{5\lambda}h, \ \ \ \ \ \
\dot{\varphi}=-\frac{2}{5\lambda}\dot{h}
\end{eqnarray}
where C is an arbitrary normalisation constant.

It's only at very late times, $z\le\sim2$, that the coupled matter establishes itself as the dominant effect on growth. This is when
we would expect the coupling's signature to start to be seen.

So far only pure curvature (adiabatic) perturbations have been considered, however isocurvature perturbations might also
exist in quintessence models \cite{abra}. For this non-minimal model we believe that their impact is negligibly small.
 Isocurvature perturbations are known to be negligible in minimally coupled tracking quintessence models. This will also be
 so for the non-minimally coupled case early on in the radiation epoch, where the couplings effect is unimportant. When the
 field is driven off tracking, close to the transition from matter to radiation, we cannot assume this, however. During the
 period when tracking is broken, the scalar field is suppressed and $\Omega_{\phi}\sim 0$ (see fig.1). In general, the
 non-adiabatic pressure perturbation $\delta p_{non-ad}$ is given by
\begin{equation}
\frac{\delta p_{non-ad}}{\rho+p}={\cal O}(\Omega_{\phi})(\delta_{\gamma}+\delta_{\phi})
\end{equation}
Therefore, since the quintessence contribution to the total energy density is highly suppressed, the isocurvature
contributions will continue to be small away from tracking behaviour, around the transition time. It is only at very late
times, after last scattering,when $\Omega_{\phi}$ is no longer small, that the isocurvature perturbations may start to grow. For the following
discussion, therefore, we only consider adiabatic perturbations.

\subsection{Implications for matter perturbations}

An important consequence of non-minimal coupling is that, when considering the coupled matter, it is the
\textit{coupled energy density}, $f\rho_{c}$, that should be interpreted as the matter
density measured in observations, not $\rho_{c}$; an analogous case is non-minimally coupled gravity,
$f(\phi)\mathcal{R}$, in which we consider the varying gravitational field strength as the observable and not constant
Newtonian gravity, $G_{N}$. So we are interested in the effective dark matter density $\tilde{\delta}_{c}$
\begin{equation}
\tilde{\delta}_{c}=\frac{\delta(f\rho_{c})}{f\rho_{c}}=\delta_{c}+\frac{f'}{f}\varphi
\end{equation}

 An insightful way to look at the coupling's effect on perturbation growth is by looking at its effect on the dimensionless growth rate
 \begin{equation}n_{eff}=\tau\frac{\dot{\tilde{\delta_{c}}}}{\tilde{\delta_{c}}}
\end{equation}

In Fig. 5 the growth rate for one scale, $k=0.1Mpc^{-1}$, is shown for various models, in each case h=0.65, $\Omega_b$=0.053. A non-minimally coupled (NMC) model
with $\Omega_\phi=0.6$ and $\lambda=8 (\beta=8, \phi_0=32.5)$, is compared with a $\Lambda$CDM model, $\Omega_\Lambda=0.6$, a sCDM model $\Omega_c=0.947$, and an analogous
minimally coupled (MC) quintessence model using the potential developed by Albrecht and Skordis \cite{albsko}
$V=V_{0}e^{-\lambda \phi}(A+(\phi-\phi_0)^{B})$ with $\lambda=8$ (A=0.01, B=2, $\phi_0=32.5$). For $z>2$ the growth rates for the
scalar field models do not differ greatly from that in the $\Lambda$CDM model.
\begin{figure}[htb]
\begin{center}
\epsfxsize=8.2cm \epsfysize=8.2cm \epsffile{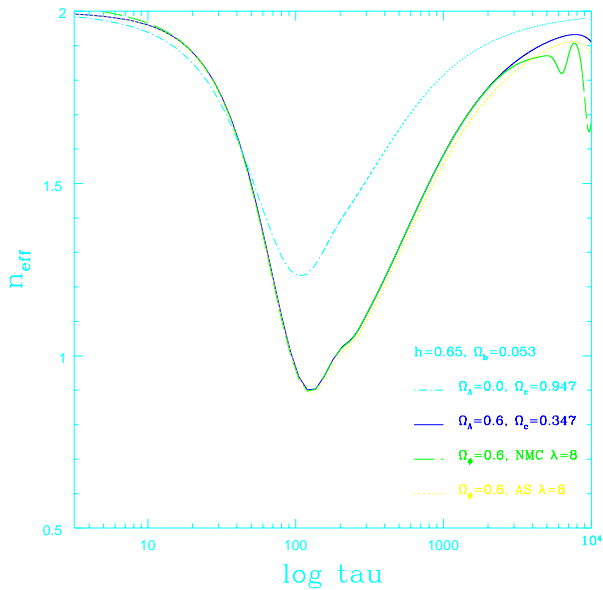} \caption{Time evolution of the effective growth
rate for 4 scenarios (all h=0.65, $\Omega_b=0.053$), 3 of which produce acceleration today: 1)$\Lambda$CDM $\Omega_\Lambda=0.6$ (full line),
2) NMC model $\Omega_\phi=0.6,\lambda=8,\beta=8, \phi_0=32.5$ (long dash) , 3) MC model $\Omega_\phi=0.6, \lambda=8, A=0.01, B=2, \phi_0=32.5$ (short
dash), and one which doesn't: 4)sCDM model $\Omega_c$=0.947 (dot-dash)}
\end{center}
\label{neff}
\end{figure}

The addition of a scalar field or cosmological constant, with $\Omega_0=1$ fixed, will act to reduce $\Omega_c$ and
therefore the size of the mass term in equation (\ref{pert}). This is the main factor responsible for the suppression of
growth at later times, rather than the non-clumping behaviour of the scalar field commonly cited as the cause.
Sub-horizon scalar field perturbations have oscillatory time evolution with decaying amplitudes, their contribution to the
evolution of matter perturbations therefore is small for the observationally interesting scales. For NMC models, the
coupling suppresses $\Omega_\phi$ around $z_{eq}$, making the scalar field contribution to $\delta_{c}$ growth
negligible. Subsequently, the growth rate for NMC models is closer to that created by a cosmological constant than for the MC
models.

At late times however, for $z<2$, the coupling and scalar field become important, and act to suppress the growth in
$\delta_{c}$ to a far greater extent than $\Lambda$ and MC models, offering a potential way to distinguish non-minimal
from minimal the NMC model.

\begin{figure}[htb]
\begin{center}
\epsfxsize=8.2cm \epsfysize=8.2cm \epsffile{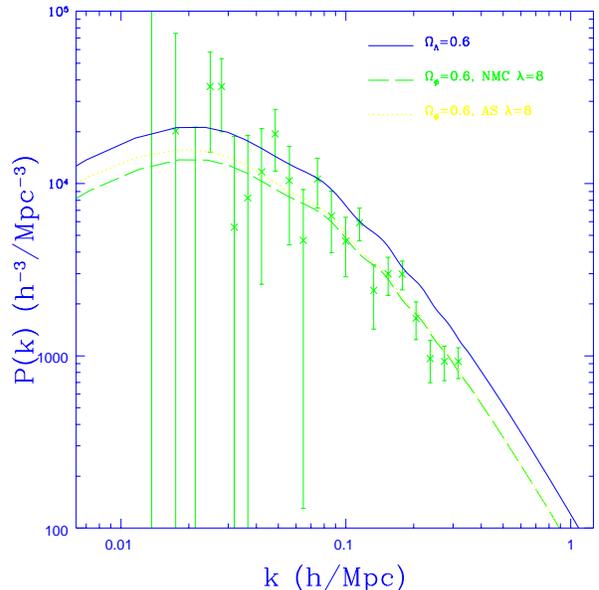} \caption{Matter power spectrum for the 3
scenarios  in Fig. 5 which produce acceleration today: $\Lambda$CDM (full), NMC (long dash) and MC (short dash), together with the de-correlated data points of Hamilton et al. Parameters in the 3 models are the same as in Fig.5.}
\end{center}
\label{matpow}
\end{figure}

The dampening effect can be also seen in the matter power spectrum P(k),
\begin{equation}
P(k)=\langle|\tilde{\delta}_{c}(k)|^{2}\rangle =(100C)8\pi^{3}h^{3}k\left(\frac{k}{k_{0}}\right)^{n-1}
\end{equation}
where C is the normalisation factor from CMBFAST \cite{CMBFAST} arising form the Bunn \& White normalisation
\cite{bunnwhite} at l=10 multipole, k is in units of h/Mpc and $k_{0}=0.05Mpc^{-1}$ and n is the tilt, chosen here
n=1 for a scale invariant spectrum.

In fig 6. the matter power spectra for the both the NMC and MC models mimic a $\Lambda$CDM model for scales $k<\sim 0.1$
i.e. those modes having entered the horizon before and around equality. There is a slight suppression but a bias factor
could in theory resolve the discrepancy. Certainly all three models give reasonable predictions for matter fluctuations
over a sphere of size $8h^{-1} {\rm Mpc}$ with $\sigma_8=$0.89,0.91,1.13 for NMC, MC and $\Lambda$CDM models
respectively, in comparison to the observed value $\sigma_8= 0.56\Omega_m^{-0.47}\sim0.9$ \cite{viana}.

For larger scales,however, the coupling does make a difference.
In scales that have only entered the horizon in recent times, whilst the coupling is important, we see a distinctive
reduction of power in comparison to the MC and $\Lambda$CDM models which tend to similar behaviour.
Although the suppression clearly distinguishes the coupled model, its profile is still consistent with current
observational results \cite{ham}. There may be an opportunity with the future SLOAN galaxy survey results to constrain
the power spectrum at these larger scales (to $k\sim 0.01$).

Another potential impact of the late time importance of the coupling is that it will affect small scale features at
$z\sim 2$, observable potentially through future weak lensing (see e.g.\cite{hoes} and references therein)and damped
Lyman $\alpha$ cloud measurements (see e.g \cite{mabert2}).

\subsection{Impact on CMB anisotropies}
 \begin{figure}[htb]
\begin{center}
\epsfxsize=8.7cm \epsfysize=8.7cm \epsffile{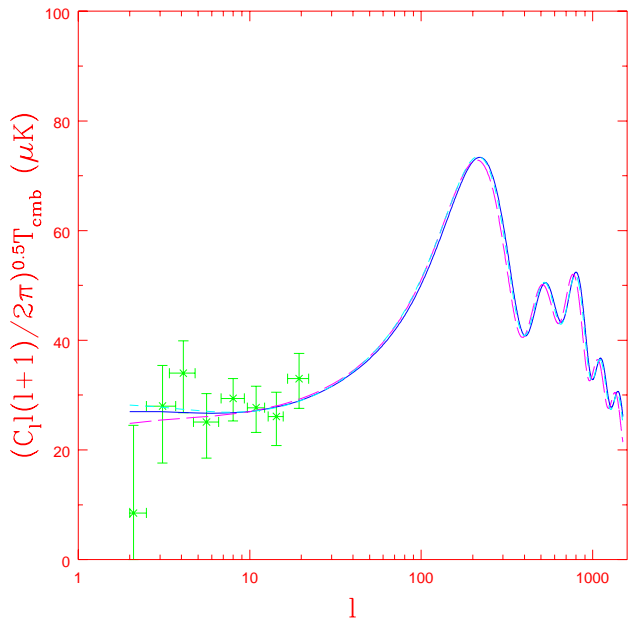} 
\caption{CMB power spectra showing low l (plateau) behaviour for the 3 scenarios in Fig. 5: $\Lambda$CDM (full), NMC (long dash) and MC (short dash) with COBE datapoints. Model parameters are the same as in Fig. 5. The 3 models evolve differently at late times producing slightly different ISW anisotropies shown in the plateau at low l. However observations at this scale are dominated by cosmic variance, so that the differences would not be observable.} 
\end{center}
\label{cmb1}
\end{figure}
Introducing a scalar field can potentially have several effects on the CMB power spectrum. Firstly, as we have already
mentioned in section ~\ref{lineq}, the scalar field gives rise to extra mass and source terms in the linear evolution
equations for $\varphi$ and $\delta_c$. These then indirectly affect the radiation perturbations,  altering the
 acoustic peak positions and heights at the time of last scattering ($\tau_{lss}$). However, the scalar
 perturbations are effectively negligible around $z_{eq}$, especially in the NMC scenario so this effect will be minimal.

Secondly, the time varying Newtonian potential after decoupling will be affected by the coupling, altering the anisotropies
produced at large angular scales (the Integrated Sachs-Wolfe effect). This can be seen in fig. 7 where the NMC model has a
different profile at small l from the MC and $\Lambda$CDM models. However the effect is not large enough to be disentangled from the effect of cosmic variance.

\begin{figure}[htb]
\begin{center}
\epsfxsize=8.7cm \epsfysize=8.7cm \epsffile{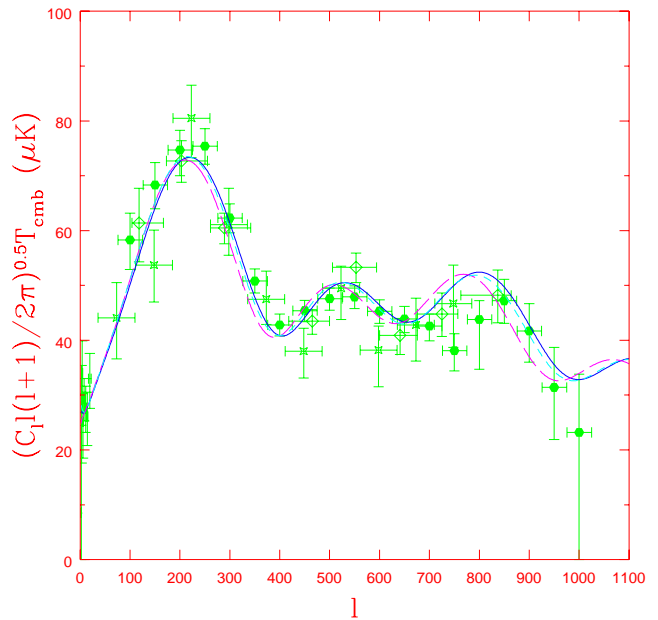} \caption{CMB power spectra showing acoustic
peaks for the 3 scenarios in fig.5:$\Lambda$CDM (full), NMC (dash) and MC (short dash), together with the data from Boomerang (solid circles), Maxima (crosses) and DASI(open diamonds).Model parameters are the same as in Fig. 5.}
\end{center}
\label{cmb2}
\end{figure}
Thirdly, the inclusion of the scalar field alters the composition of the energy density, altering the angular diameter
distance of the acoustic horizon size at recombination. This can be parameterised by the value of ${\cal B}$
\begin{equation}
{\cal B}=\Omega_c^{1/2}h\int_{z_{rec}}^{z_{0}}dz\left\{\Sigma\Omega_{j}z^{3(1+w_{j})}\right\}^{-1/2}
\end{equation}
Altering the value of ${\cal B}$ shifts the positions of the peaks. The critical problem one confronts when trying to use
CMB spectra to differentiate between models is the degeneracy that exists between models with identical $\Omega_c,\Omega_b$
and ${\cal B}$ \cite{efbond}. It has been shown that this degeneracy can be broken for scalar field models in which a large
fraction of the energy density at $\tau_{lss}$ is from the scalar \cite{BHM}; the scalar field acting as an effective increase in the number of relativistic degrees of freedom. However for models in which $\lambda \ge8$
the degeneracy still exists in minimally coupled models. In Figs. 7 and 8, CMB spectra are plotted for the scenarios discussed in the previous section against COBE \cite{COBE}, MAXIMA \cite{MAX}, Boomerang \cite{Boom} and DASI \cite{DASI} data.  All the models discussed have ${\cal B}$=1.77 and yet one can see that the degeneracy of the first peak is slightly broken, with the NMC model having $l_{peak}$=215 in comparison to 224 for both the MC and $\Lambda$CDM models. It is also interesting to note that the CMB spectra for coupled models with different $\lambda$ values are effectively degenerate in themselves, as shown in the figure. This implies that, although coupling itself may be distinctive, CMB spectra will not be able to isolate the parameter in the potential.

The fourth possible effect is on the separation of the peaks. This has been proposed as a possible mechanism with which to
distinguish minimally coupled models \cite{doran}. They are not distinguishable from $\Lambda$CDM if
$\Omega_\phi(\tau_{lss})$ is small however, as mentioned above. But in the case of non-minimally coupled models the
degeneracy in the second and third peaks \textit{is} broken because of the effect the coupling has on $\tau_0$ the
conformal time nowadays. This is of particular interest given the expected improvements in peak definition
(e.g  \cite{Boom},\cite{MAX},\cite{DASI},\cite{CBI}) .
 The separation of the peaks $dl$ is given by
\begin{equation}
\delta l=\pi \frac{\tau_{0}-\tau_{lss}}{r_{s}}
\end{equation}
where $r_s$ is the sound horizon and $c_{s}$, the baryon speed of sound, both of which can be assumed effectively constant
across the models. The NMC model has $\tau_0$=12,530 in comparison to  $\tau_0$=13,077 for the $\Lambda$CDM model. This
reduces the separation slightly breaking the degeneracy, as shown; the separation of the first and second peaks in the NMC model is dl=309 in comparison to 327 $\Lambda$CDM scenario.

Although distinguishable from the cosmological constant spectrum, the
difference is still too small to be resolved with current observational data, including the most recent Boomerang \cite{Boom} and DASI \cite{DASI} data, showing highly improved definition in the second and third peaks. However with a number of observational projects continuing to focus attention on resolving the higher peaks, the breaking of
 degeneracy may offer a way to constrain non-minimally coupled models. In Fig.9 we plot the residual differences between the $\Lambda$ CDM and NMC $C_l$ spectra in Fig.8 when compared with estimated MAP errors. The parameters used to estimate the MAP errors are shown in appendix \ref{APP3}. The estimated errors are considerably smaller than these residual levels for $l<900$ implying that we may be able to distinguish between these various models within the near future.
\begin{figure}[htb]
\begin{center}
\epsfxsize=8.7cm \epsfysize=8.7cm \epsffile{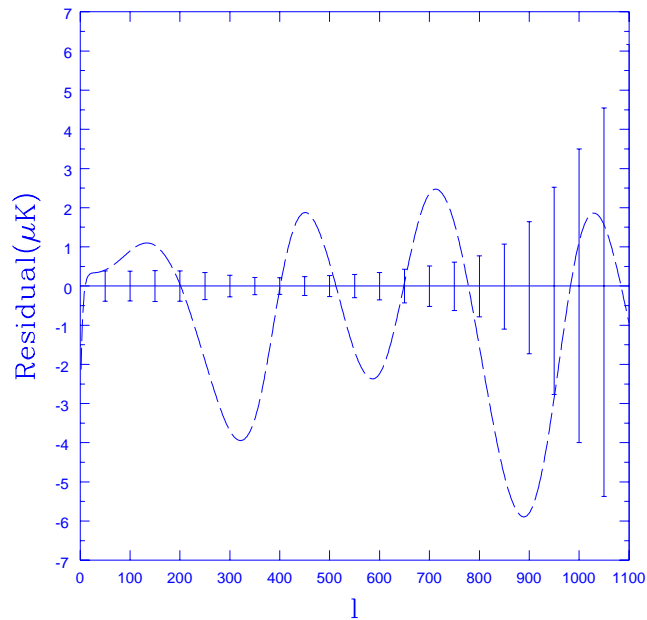} \caption{Residual deviation of the the NMC model's temperature fluctuations from those of the $\Lambda$CDM model, both shown in Fig.8., with estimated MAP error bars.}
\end{center}
\label{resid}
\end{figure}
\section{Conclusion}
We have examined the impact a scalar field, non-minimally coupled to cold dark matter will have on the evolution of matter
and radiation perturbations. We considered firstly its impact on the linear evolution equations and found that even though
it did introduce new terms these were effectively negligible for all but very late times. The impact of this late time
behaviour was then considered for matter perturbations where it was seen to create a suppression of growth at large scales.
The coupling was also found to break the degeneracy usually seen in the CMB spectrum, slightly shifting the position of the first peak and reducing the separation between adjacent peaks. These two distinctive `signatures' of the coupled dark energy model are not  resolvable with current
observations.  However projects currently underway look to mapping both the matter power spectrum and CMB peaks with much improved accuracy. These may
offer an opportunity to eventually distinguish between $\Lambda$CDM, minimally coupled and non-minimal coupled quintessence models in the near future.

\section{Acknowledgements}
I would like to thank Jo\~{a}o Magueijo and Carlo Contaldi for many helpful conversations. RB acknowledges the financial support from PPARC.
\section{Appendix}
\subsection{Bianchi's Identity}\label{APP1}
In this section we derive the equations of motion for the coupled dark matter and scalar fields explicitly from the action.
\begin{equation}
{\cal S}=\int \sqrt{-g} d^{4}x {\left( -{{\cal R}\over 2}
+{1\over 2}\phi_{,\mu}\phi^{,\mu}-V(\phi)
+{\cal L}_V+f(\phi){\cal L}_{c} \right)}
\end{equation}
Bianchi's identity reflects the symmetry of the Riemann tensor, the Einstein tensor being convariantly conserved,
\begin{equation}\label{B1}
({\cal R}^{\mu\nu}-{1 \over 2}g^{\mu\nu}{\cal R})_{;\nu}=(T^{\mu\nu(\phi)}+T^{\mu\nu (V)}+f(\phi)T^{\mu\nu(c)})_{;\nu}=0
\end{equation}
Since visible matter is minimally coupled in the model we can immediately separate it out,
\begin{equation}\label{B2}
T^{\mu\nu(V)}_{;\nu}=0
\end{equation}
Using the explicit definition of the energy momentum tensor in terms of their Lagrangian,
\begin{eqnarray}
T^{\mu\nu}_{;\nu}&=&\left[\frac{2}{\sqrt{-g}}\left({\partial(\sqrt{-g}{\cal{L}}) \over \partial g_{\mu\nu}}
\right)\right]_{;\nu}\label{TLrel}
\end{eqnarray}
so the scalar field component of the Bianchi identity in terms of $\phi$ and its derivatives is
\begin{eqnarray}
T^{\mu\nu (\phi)}_{;\nu}=
\phi^{,\mu}\left(\phi^{,\nu}_{,\nu}+\Gamma^{\nu}_{\alpha \nu}\phi^{,\alpha}+V'(\phi)\right)
\label{E0}
\end{eqnarray}
The Euler-Langrange equation, which is just the Klein Gordon equation, allows us to simplify this further
\begin{eqnarray}
\phi^{,\alpha}_{,\alpha}+\Gamma^{\beta}_{\alpha\beta}\phi^{,\alpha}+V'(\phi)=f'(\phi){\cal L}_{c}
\label{E1}
\end{eqnarray}
Combining these two expressions we obtain
\begin{eqnarray}
 T^{\mu\nu (\phi)}_{;\nu}=f'(\phi){\cal L}_{c}g^{\mu\nu}\phi_{,\nu} \label{E3}
\end{eqnarray}
For the coupled matter then,
\begin{eqnarray}
\left(f(\phi)T^{\mu\nu (c)}\right)_{;\nu}
&=&  f' \phi_{;\nu}T^{\mu\nu (c)}+fT^{\mu\nu (c)}_{;\nu} \label{E2}
\end{eqnarray}
Combining the results in equations (\ref{B2}) and (\ref{E3}), Bianchi's identity in (\ref{B1}) is given by
\begin{eqnarray}
T^{\mu\nu (c)}_{;\nu}={f' \over f}\phi_{;\nu}\left({\cal L}_{c}g^{\mu\nu}-T^{\mu\nu (c)}\right)\label{biam}
\end{eqnarray}

We can obtain an expression for the Lagrangian for perfect fluid by considering it to be a gas of particles with masses $m_{a}$  and paths $x^{i}_{a}$ \cite{Peebles}
\begin{eqnarray} {\cal L}_{a}(x)&=&-
 m_{a}\delta(x-x_{a}(t))\left(-g_{\mu\nu}\dot{x}^{\mu}\dot{x}^{\nu}\right)^{1\over 2}\label{lag}\end{eqnarray}
By noting that the length of the 4-velocity $\left(-g_{\mu\nu}{dx^{\mu}\over d\lambda}{dx^{\nu}\over d\lambda}\right)^{1\over 2}=ds/d\lambda$ equals 1 for dust, this expression simplifies greatly. Averaging over particles in the gas rest frame we find ${\cal L}_{c}=-n\langle m/u^{0} \rangle$ where n is the particle number density and $u^{\mu}=dx^{\mu}/d\lambda$ is the 4-velocity. For pressureless particles $u^{\mu}=\{1,0,0,0\}$ therefore ${\cal L}_{c}=-\rho_{c}$. 

We can also obtain an expression for the stress-energy tensor from the Lagrangian in an analogous way.
Using the relationship between $T^{\mu\nu}$ and ${\cal L}$ in (\ref{TLrel}) the stress-energy for a particle is given by
\begin{eqnarray}T^{\mu\nu}_{a}= m_{a}{\delta(x-x_{a}(t))\over \sqrt{-g}}\dot{x}^{\mu}\dot{x}^{\nu}u^{0}\end{eqnarray}
On averaging over the particles the energy density is $\rho=\langle T^{00}\rangle=n\langle mu^{0}\rangle$ and the pressure in the $x$-direction is given by $p=\langle T^{11}\rangle=n\langle mu^{0}(v^{1})^{2}\rangle$ where $v^{i}=dx^{i}/dt=u^{i}/u^{0}$, for dust therefore $p_{c}=0$. In addition, in the rest frame there is zero streaming velocity so that $\langle T^{0i}\rangle=n\langle mu^{0}v^{i}\rangle=0$. 

In this paper we assume that the coupled matter is comprised of cold pressureless dust particles. Putting the expression for ${\cal L}_{c}$ into (\ref{biam}), the background evolution equation for non-minimally coupled matter is identical to that in the minimally coupled case.
\begin{eqnarray}
T^{\mu\nu (c)}_{;\nu}=\dot{\rho}_{c}+3H\rho_{c}=0 
\end{eqnarray} 

\subsection{Linear perturbation equations}\label{APP2}
We here derive the linear perturbation equations for the coupled cdm and scalar fields in detail.We assume the notation of Ma and Bertschinger \cite{mabert} and results of Ferreira and Joyce \cite{ferjoy}.
Consider perturbations to a flat FRW metric in the synchronous gauge, with line element
\begin{eqnarray}
ds^{2}=a(\tau)^{2}\left\{-d\tau^{2}+(\delta_{ij}+h_{ij})dx^{i}dx^{j}\right\} \nonumber 
\end{eqnarray}
By considering the perturbed Euler-Lagrange equation for the scalar field we can obtain the linear
 evolution equation for $\varphi$.
 \begin{eqnarray}
\varphi^{,\alpha}_{,\alpha}+\delta\Gamma^{\beta}_{\alpha\beta}\phi^{,\alpha}
+\Gamma^{\beta}_{\alpha\beta}\varphi^{,\alpha}+V''\varphi=f''\varphi{\cal L}_{c}+f'\delta {\cal
L}_{c}\label{S2a} \end{eqnarray}
Perturbing the particle Lagrangian in (\ref{lag})
\begin{eqnarray} \delta{\cal L}_{a}&=&-
 \delta m_{a} \delta(x-x_{a}(t))\left(-g_{\mu\nu}\dot{x}^{\mu}\dot{x}^{\nu}\right)^{1\over 2}\label{plag}\end{eqnarray}
and averaging over all dust particles, we have $\delta {\cal L}=-n\langle \delta m/u^{0} \rangle=-\delta\rho_{c}$.
The non-minimally coupled scalar perturbation equation becomes
\begin{eqnarray} {\ddot{\varphi}\over a^{2}}+{1\over 2}\dot{h}\dot{\phi}+2{\dot{a}\over
a}\dot{\varphi}+V''\varphi-\varphi^{,i}_{,i}=-\rho_{c}(f''\varphi+f'\delta_{c})
\label{S2b}\end{eqnarray}
The perturbed Einstein equations can be used to obtain an expression for $\dot{\delta\rho}$
\begin{eqnarray}
(\rho+P)\theta&+&3H(\delta\rho+\delta P)+{1\over 2}(\rho+P)\dot{h}+\dot{\delta\rho}=0\label{PE6}
\end{eqnarray}
We are interested in the interacting dark matter and scalar field, which are not separable in (\ref{PE6})
\begin{eqnarray}
f\rho_{c}\theta_{c}&+&3H\delta(f\rho_{c})+{1\over 2}(f\rho_{c})\dot{h}+{d\over d{\tau}}(\delta(f\rho_{c}))\nonumber \\
&+&\dot{\phi}{\nabla^{2}\varphi\over a^{2}}+3H\left({2\dot{\phi}\dot{\varphi}\over a^{2}}\right)+{1 \over 2}{\dot{\phi}^{2}\over a^{2}}a^{2}\dot{h}\nonumber \\ &+&\left({\ddot{\phi}\dot{\varphi}\over a^{2}}+{\dot{\phi}\ddot{\varphi}\over a^{2}}+2{\dot{a}\over a}{\dot{\phi}\dot{\varphi}\over a^{2}}V'\dot{\varphi}+V''\varphi\dot{\phi}\right)
=0 \label{comb}
\end{eqnarray}
Using the equations of motion for the background and perturbed scalar field we find that (\ref{comb}) simplifies substantially. Interestingly, we find that the coupling does not affect the first order dark matter perturbation equation
\begin{eqnarray}
\dot{\delta_{c}}=-\theta_{c}-{1\over 2}\dot{h}&& \label{M2}
\end{eqnarray}
With the residual degree of freedom in the synchronous gauge we are free to fix one additional
parameter, by convention we set $\theta_{c}=0$ so that $\dot{\delta_{c}}={1 \over 2}\dot{h}$. Ignoring baryons, for simplicity, the second order perturbation equation becomes,
\begin{eqnarray}
\ddot{\delta}_{c}&+&H\dot{\delta}_{c}+{3H^{2}f\Omega_{c}\over 2}\delta_{c}= \nonumber \\&& -3H^{2}\Omega_{\gamma}\delta_{\gamma}-2\dot{\phi}\dot{\varphi}+(a^{2}V'-{3H^{2}\Omega_{c}f'\over 2})\varphi 
\end{eqnarray}

\subsection{MAP error bar estimation}\label{APP3}
The standard error on the estimate of $C_{l}$, $\Delta C_{l}$ for an experiment with $N$ frequency channels (denoted subscript n), each respectively with angular resolution $\theta_{n,fwhm}(arcmin)=\theta_{n}(rad)/60\times\pi/180$ and sensitivity $\sigma_{n}$ per resolution element, scanning a fraction $f_{sky}$ of the sky in bins of $l$ size $\Delta l$ is given by \cite{BET97}
\begin{eqnarray}
\Delta C_{l}&\approx&\left({2\over (2l+1)f_{sky}\Delta l}\right)^{0.5}\left[C_{l}+\bar{\omega}^{-1}\bar{{\cal B}}_{l}^{-2}\right]\nonumber \\ \bar{\omega}&\equiv&\sum_{n}\omega_{n},\ \ \ \ \bar{{\cal B}}_{l}^{2}\equiv\bar{\omega}^{-1}\sum_{n}{\cal B}_{nl}^{2}\omega_{n}, \nonumber \\ \omega_{n}&\equiv&(\sigma_{n}\theta_{n})^{-2},\ \ \ \ {\cal B}_{nl}^{2}\approx e^{-l(l+1)/l_{s}^{2}}
\end{eqnarray}
where we have assumed that the experimental beam is approximately Gaussian filtering scale $l_{s}\equiv\sqrt{8ln2}\theta_{n}^{-1}$.
We have assumed a useful sky fraction $f_{sky}=0.65$ and $\Delta l=50$, Table 1 shows the remaining parameters used, taken from \cite{Teg2}.
\begin{table}[h]
\caption{MAP CMB Experimental Specifications}
\begin{center}
\footnotesize
\begin{tabular}{c c c c}
 \ \ $\nu$ (GhZ) \ \ & \ \ $\theta_{n,fwhm}$ \ \ & \ \ $10^{6}\sigma$ \ \ & \ \ $N_{ch}$\ \ \\
\hline
40 & 28' & 8.2 & 4 \\
60 & 21' & 11.0 & 4 \\
90 & 13' & 18.3 & 8 \\ \hline 
\end{tabular}
\end{center}
\end{table}

The combined errors in Fig.9 are therefore $\Delta [C_{l}(\Lambda CDM)-C_{l}(NMC)]$=$\Delta C_{l}(\Lambda CDM)$ +$ \Delta C_{l}(NMC)$.

\end{document}